\documentclass{iau}
\usepackage{graphicx}
\usepackage{amsmath}
\usepackage{amsfonts}
\usepackage{amssymb}
\usepackage{color}

\usepackage{hyperref}

\definecolor{green2}{rgb}{0,0.61,0}
\hypersetup{
   colorlinks,%
  citecolor=green2,%
  filecolor=blue,%
  linkcolor=red,%
  urlcolor=blue
 }

\title[Semi-analytical study on the generic degeneracy for galaxy clustering ... ] 
{Semi-analytical study on the generic degeneracy for galaxy clustering measurements}

\author[Alejandro Guarnizo, Luca Amendola, Martin Kunz \& Adrian Vollmer ]   
{Alejandro Guarnizo$^1$, Luca Amendola$^1$, Martin Kunz$^2$ \and Adrian Vollmer$^1$ }

\affiliation{$^1$Institut F\"{u}r Theoretische Physik, Ruprecht-Karls-Universit\"{a}t Heidelberg,
Philosophenweg 16, 69120 Heidelberg, Germany \\[\affilskip]
$^2$D\'epartement de Physique Th\'eorique and Center for Astroparticle Physics,
Universit\'e de Gen\`eve, Quai E.\ Ansermet 24, CH-1211 Gen\`eve 4, Switzerland
}

\pubyear{2014}
\volume{306}  
\pagerange{119--126}
\jname{Statistical Challenges in 21st Century Cosmology}
\editors{A. F. Heavens, J.-L. Starck \& A. Krone-Martins, eds.}
\begin{document}

\maketitle

\begin{abstract}
From the galaxy power spectrum in redshift space, 
we derive semi-analytical results on the generic degeneracy of galaxy clustering measurements. Defining the
observables $\bar{A}= Gb\sigma_8$ and $\bar{R} = Gf\sigma_8$, (being $G$ the growth function, 
$b$ the bias, $f$ the growth rate, and $\sigma_8$ the amplitude of the power spectrum), we perform a Fisher matrix analysis to
forecast the expected precision of these quantities for a Euclid-like survey. Among the results
we found that galaxy surveys have generically a slightly negative correlation between $\bar{A}$ and $\bar{R}$,
and they can always measure $\bar{R}$ about 3.7 to 4.7 times better than $\bar{A}$.
\end{abstract}

\firstsection 
\section{Introduction}

Future galaxy surveys will provide new opportunities to verify the current standard cosmological
model, and also to constrain modified gravity theories, invoked to explain the present accelerated
expansion of the universe. Before studying general parametrizations of dark energy, its however important to understand 
first which quantities can be really observed.
From this direction recently \cite{Amendola2013} shown 
that cosmological measurements can determine, in addition to the expansion
rate $H(z)$, only three additional variables $R$, $A$ and $L$,
given by 
\begin{equation}\label{eq:DirectObs}
A  =Gb\delta_{\text{m,0}}\,,\qquad R=Gf\delta_{\text{m,0}}\, , \qquad
L  =\Omega_{\text{m}0}GY(1+\eta)\delta_{\text{m,0}}\,. 
\end{equation}
with $G$ is the growth function,
$b$ is the galaxy bias with respect to the dark matter density
contrast, and $\delta_{\text{m,0}}$ is the dark matter density
contrast today. The functions $\eta$ (the anisotropic stress $\eta = -\frac{\Phi}{\Psi}$ ) and $Y$ 
(the clustering of dark energy  $Y\equiv-\frac{2k^{2}\Psi}{3\Omega_{\text{m}}\delta_{\text{m}}} $), 
describe the impact of the dark energy on cosmological perturbations. In \cite{Amendola2014}, a Fisher analysis was made using
galaxy clustering, weak lensing and supernovae probes, in order to find the expected accuracy with which an Euclid-like survey 
can measure the anisotropic stress $\eta$, in a model-independent way. \\

In this work we want to obtain some results on the intrinsic degeneracy on galaxy clustering measurements, using the quantities 
 $A$ and $R$. We  use a flat $\Lambda$CDM fiducial model, with $\Omega_{\text{m},0}h^{2}=0.134$, $\Omega_{b,0}h^{2}=0.022$, $n_{s}=0.96$,
$\tau=0.085$, $h=0.694$, $\Omega_{k}=0$, Euclid-like survey specifications are used  \cite{Amendola2013-2}: 
we divided the redshift range $[0.5,2.0]$ in 5 bins of width $\Delta z = 0.2$ and one of width $\Delta z = 0.4$;
a spectroscopic error $\delta z= 0.001(1+z)$, and a fraction of sky $f_{\text{sky}} = 0.375$; the bias $b$ in the fiducial is assumed to be unity.

\section{Fisher matrix for Galaxy Clustering}

Observations of the growth rate $f$ from large scale structures 
using Redshift Space Distortions (RSD), give a direct way to test different dark energy models, \cite{Song2009}, \cite{Percival2009}, \cite{Racanelli2013}. 
Let us consider now the galaxy power spectrum in redshift space
 \begin{equation}
P(k,\mu)=(A+R\mu^{2})^{2}=(\bar{A}+\bar{R}\mu^{2})^{2}\delta_{\text{t,0}}^{2}(k),\label{gcpower}
\end{equation}
whit $\bar{A}=Gb\sigma_{8},\bar{R}=Gf\sigma_{8}$, and we explicitly use $\delta_{\text{m,0}}=\sigma_8 \delta_{\text{t,0}}$. The Fisher matrix is in general
\begin{equation}
F_{\alpha\beta}=\frac{1}{8\pi^{2}}\int_{-1}^{1}d\mu\int_{k_\text{min}}^{k_\text{max}}  k^{2} V_{\text{eff}}D_{\alpha}D_{\beta}\, {\mathrm d k},
\end{equation}
 where $D_{\alpha}\equiv\frac{d\log P}{dp_{\alpha}}$,  and $V_{\text{eff}}$ is the effective volume of the survey
\begin{equation}\label{veff}
V_{\text{eff}} = \left(\frac{\bar{n} P(k,\mu)}{\bar{n} P(k,\mu) + 1 }\right)^2 V_{\text{survey}},\end{equation}
$\bar{n}$ being the galaxy number density in each bin. We want to study the dependence
on the angular integration in the Fisher matrix for the set of parameters
$p_{\alpha}=\{\bar{A}(z_{\alpha}),\bar{R}(z_{\alpha})\}$.
The derivatives of the power spectrum are 
\begin{equation}
D_{\alpha}=\frac{2}{\bar{A}+\bar{R}\mu^{2}}\{1,\mu^{2}\}.
\end{equation}
We consider two cases depending on the behavior of $V_{\text{eff}}$, equation (\ref{veff}): 
\begin{enumerate}
\item ``Enough data'' $\bar{n}P(k,\mu)\gg1$, then we have $V_{\text{eff}}\approx V_{\text{survey}}$
and the Fisher matrix could be written as 
\begin{equation}
F_{\alpha\beta}\approx\frac{1}{2\pi^{2}}\int_{k_{\text{min}}}^{k_{\text{max}}}  k^{2} V_{\text{survey}} M_{\alpha\beta}\, {\mathrm d k},\label{FishGc2}
\end{equation}
 where 
 {\small
\begin{equation}
M_{\alpha\beta}=4\left(\begin{array}{cc}
\dfrac{\sqrt{\bar{S}}+(\bar{A}+\bar{R})\tan^{-1}\sqrt{P_{1}}}{\bar{A}^{3/2}\sqrt{R}(\bar{A}+\bar{R})} & \dfrac{-\sqrt{\bar{S}}+(\bar{A}+\bar{R})\tan^{-1}\sqrt{P_{1}}}{\bar{R}^{3/2}\sqrt{A}(\bar{A}+\bar{R})}\\
\\
\dfrac{-\sqrt{\bar{S}}+(\bar{A}+\bar{R})\tan^{-1}\sqrt{P_{1}}}{\bar{R}^{3/2}\sqrt{A}(\bar{A}+\bar{R})}\; & \;\dfrac{\bar{R}(3\bar{A}+2\bar{R})-3(\bar{A}+\bar{R})\sqrt{\bar{S}}\tan^{-1}\sqrt{P}_{1}}{\bar{R}^{3}(\bar{A}+\bar{R})}
\end{array}\right)
\end{equation}
}
being $\bar{S}=\bar{A}\bar{R}$ and $P_1=\bar{R}/\bar{A}$. \\
\item Shot-noise dominated $\bar{n}P(k,\mu)\ll1$, then $V_{\text{eff}}\approx(\bar{n}P(k,\mu))^{2}V_{\text{survey}}$
and since we are interested only in the $\mu$ dependence, we can
write $V_{\text{eff}}\approx P(k,\mu)^{2}$. Then the Fisher matrix
becomes
\begin{equation}
F_{\alpha\beta}\approx\frac{1}{2\pi^{2}}\int_{k_{\text{min}}}^{k_{\text{max}}}  k^{2}\delta_{\text{t,0}}^{4}(k) N_{\alpha\beta}\, {\mathrm d k},\label{FishGc3}
\end{equation}
 with 
{\small
\begin{equation}
N_{\alpha\beta}=8\left(\begin{array}{cc}
\bar{A}^{2}+\dfrac{2\bar{A}\bar{R}}{3}+\dfrac{\bar{R}^{2}}{5}\;\; & \;\;\dfrac{\bar{A}^{2}}{3}+\dfrac{2\bar{A}\bar{R}}{5}+\dfrac{\bar{R}^{2}}{7}\\
\\
\dfrac{\bar{A}^{2}}{3}+\dfrac{2\bar{A}\bar{R}}{5}+\dfrac{\bar{R}^{2}}{7} & \dfrac{\bar{A}^{2}}{5}+\dfrac{2\bar{A}\bar{R}}{7}+\dfrac{\bar{R}^{2}}{9}
\end{array}\right).
\end{equation}
}
 
\end{enumerate}

\begin{figure}[tb]
\begin{center}

 \includegraphics[width=0.42\hsize]{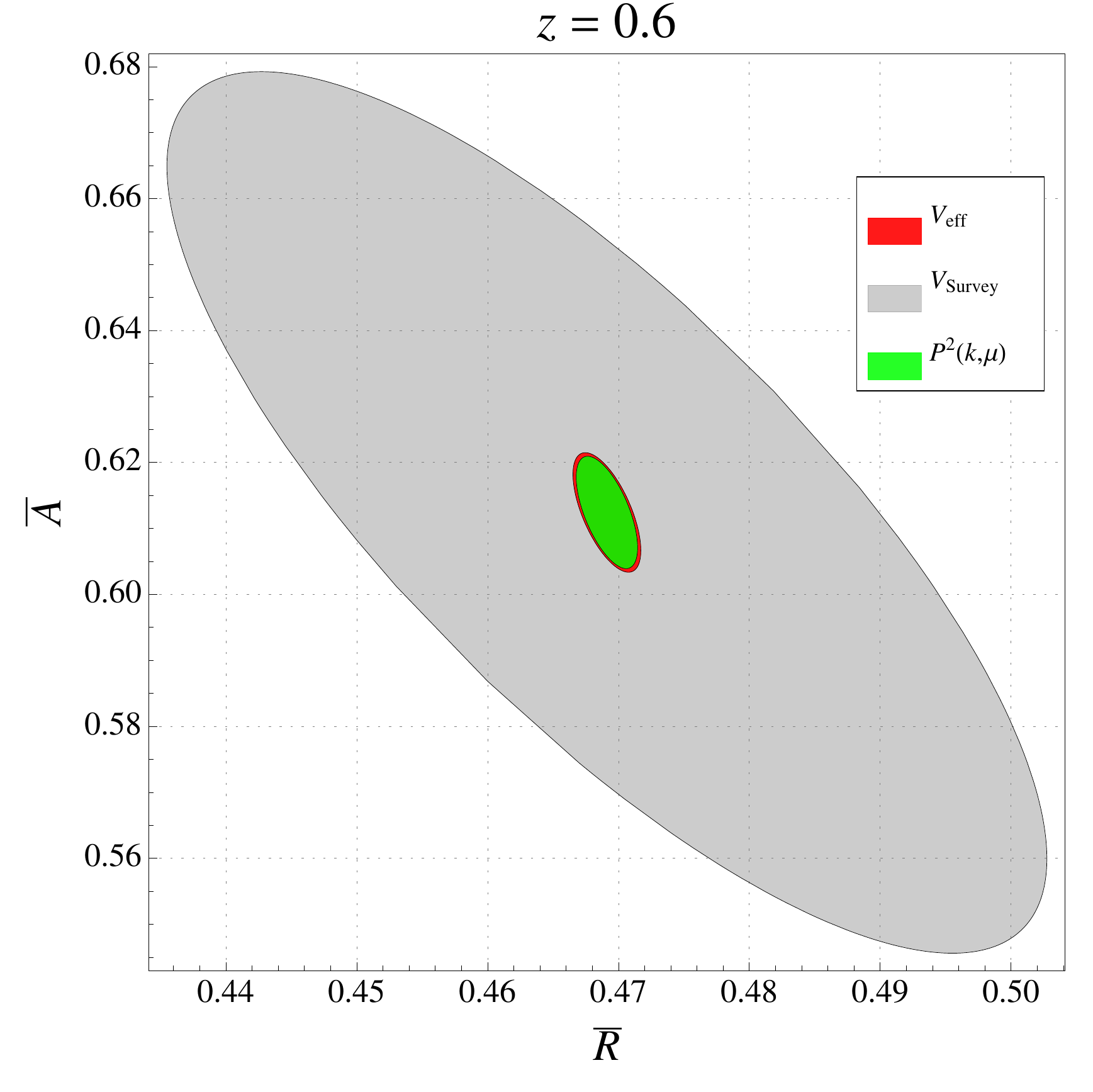}\hspace{4mm}
        \includegraphics[width=0.42\hsize]{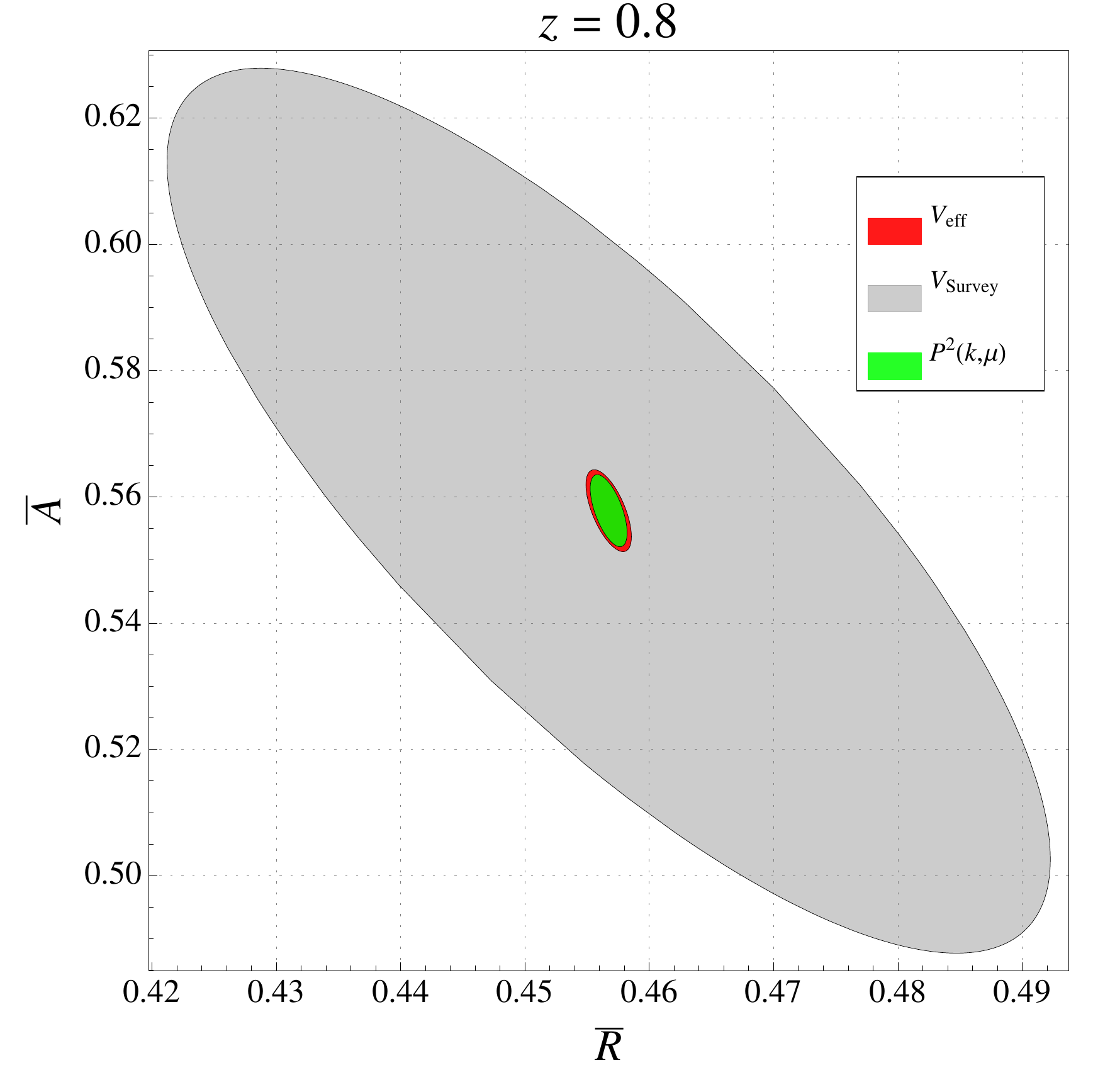}\\[4mm]
        \includegraphics[width=0.42\hsize]{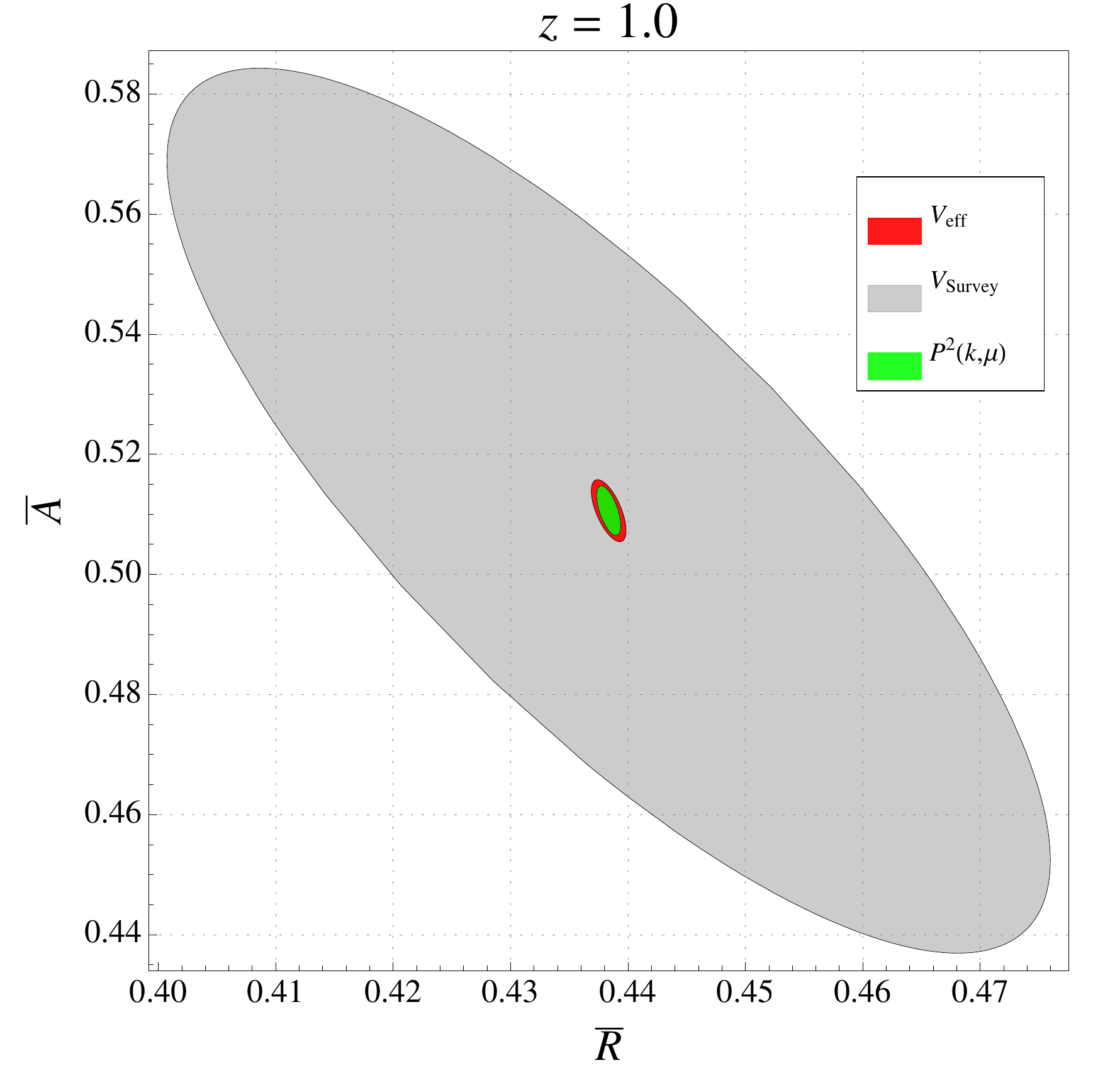}\hspace{4mm}
        \includegraphics[width=0.42\hsize]{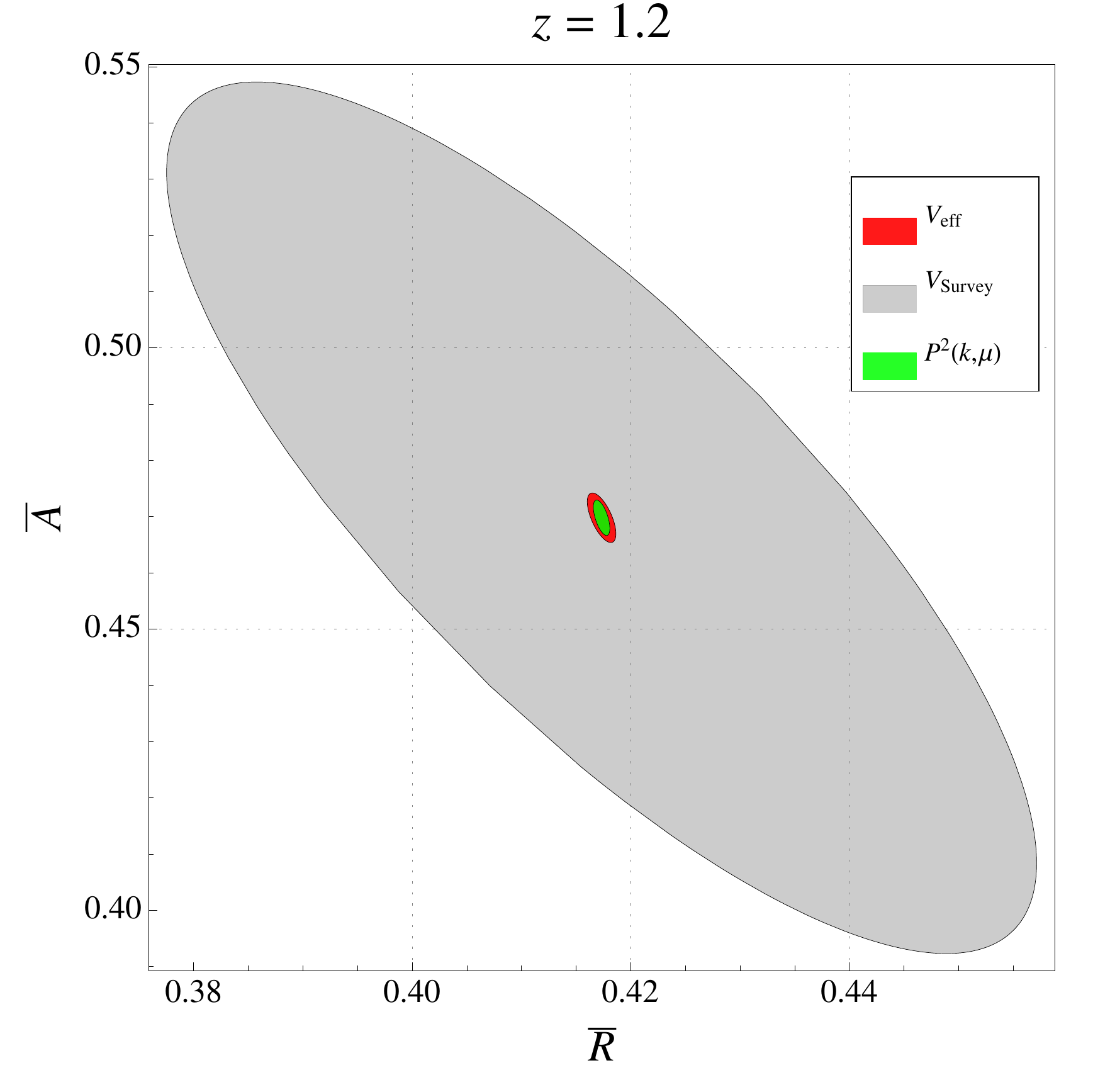}\\[4mm]
        \includegraphics[width=0.42\hsize]{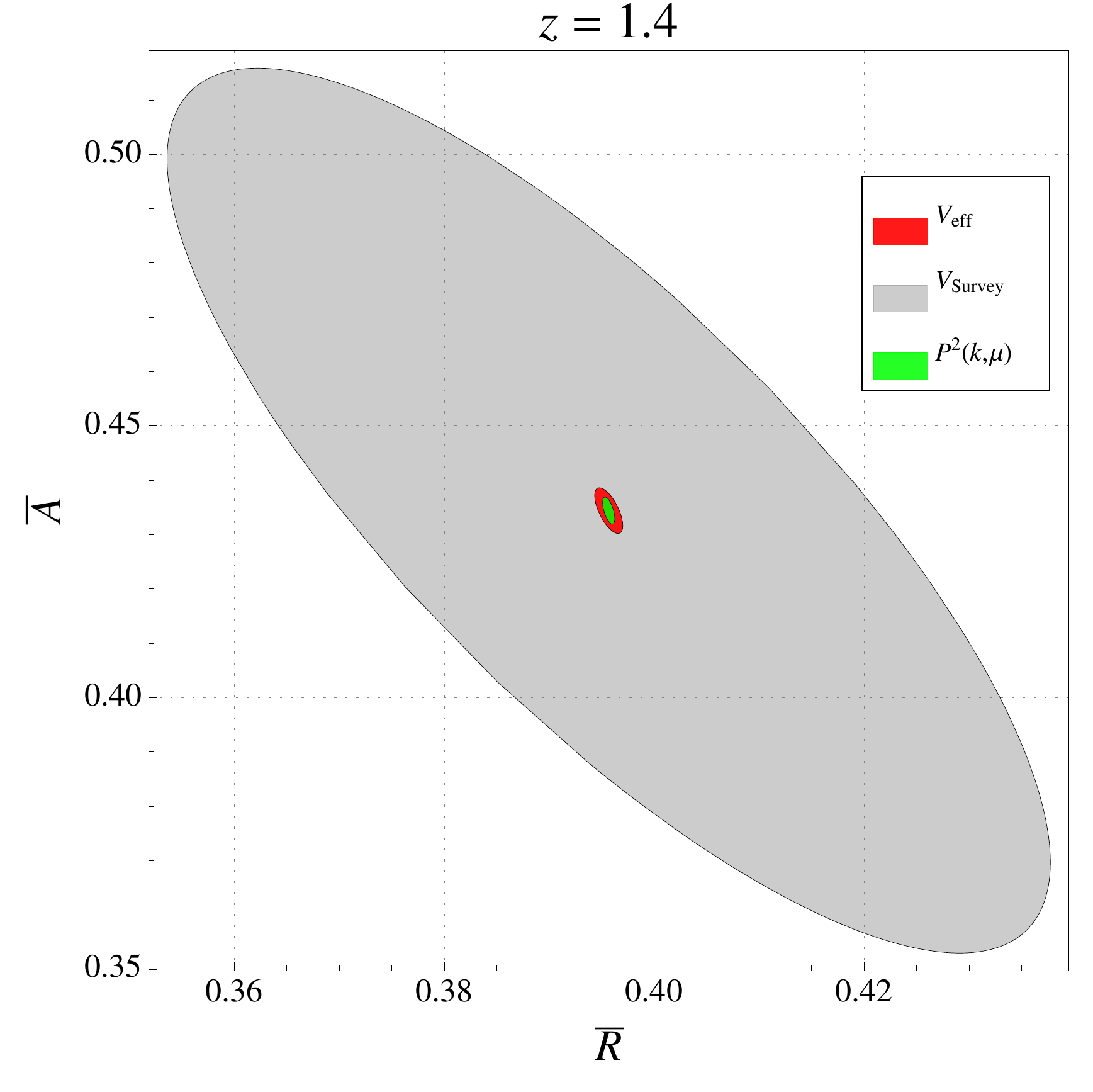}\hspace{4mm}
        \includegraphics[width=0.42\hsize]{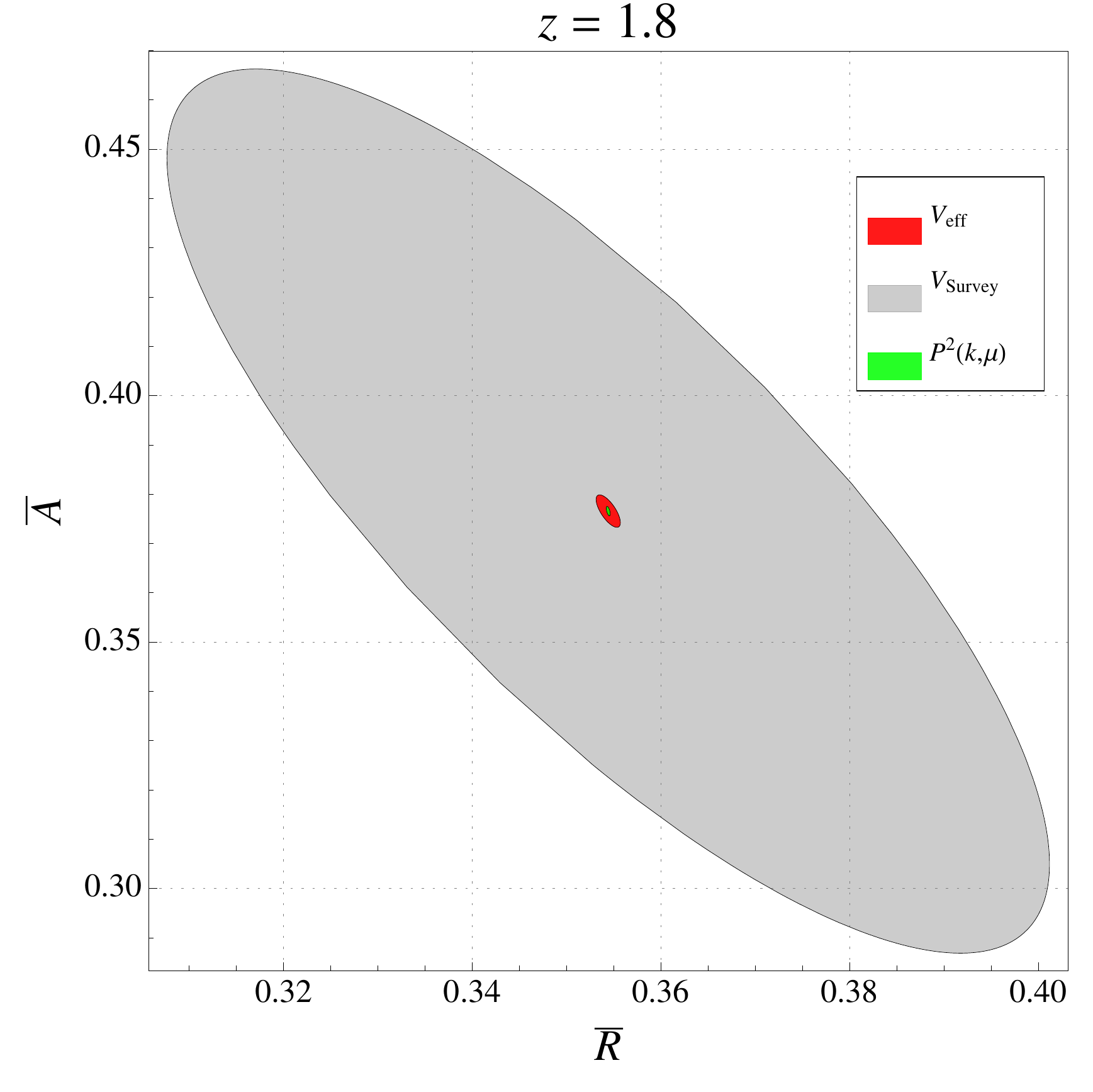}
 \caption{(color online). Confidence contours for $\bar{A}$ and $\bar{R}$ in the three cases: orange line $V_{\text{eff}}$, blue line 
         $V_{\text{eff}}\approx V_{\text{survey}}$, and green line $V_{\text{eff}}\approx P(k,\mu)^{2}$.}
   \label{fig1}
\end{center}
\end{figure}

We notice that in the two limiting cases above, we can move the matrices
$M_{\alpha\beta}$ and $N_{\alpha\beta}$ outside of the integral,
as for the fiducial model $\bar{A}$ and $\bar{R}$ do not depend
on $k$. This means that, although the absolute size of the error
ellipse depends on the integral, the relative size and orientation
do not. In other words, we can obtain `generic expectations' for the
shape of the degeneracy between $\bar{A}$ and $\bar{R}$ from galaxy
clustering surveys. These results are quite representative for the full range of $\bar A$ and $\bar R$, i.e.\ galaxy surveys
have generically a slightly negative correlation between $\bar A$ and $\bar R$, and they can always measure
$\bar R$ about 3.7 to 4.7 times better than $\bar A$, see Figure \ref{fig1}. In comparisson to the results of \cite{Song2010}, 
we remove the dependence on $\delta_{\text{t,0}}$, eq. (\ref{eq:DirectObs}), which is a quantity that depends on inflation or other
primordial effects.

\begin{acknowledgements} 
L.A., A.G. and A.V. acknowledge support
from DFG through the project TRR33 ``The Dark Universe'', A.G. also acknowledges support from DAAD through program 
``Forschungsstipendium f\"{u}r Doktoranden und Nachwuchswissenschaftler''.  M.K. acknowledges financial support from the Swiss NSF.

\end{acknowledgements}

\end{document}